\documentclass[12pt]{article}
\usepackage{epsf,latexsym,rotate}
\usepackage[ps,arc,frame]{xy}
\usepackage{amsfonts,amssymb} 
\usepackage{psfrag}
\epsfverbosetrue
\usepackage{graphicx}

\newcommand{\rxyn}[2]{{\begin{xy} 0;<2mm,0mm>:<0mm,2mm>::0;0,
,(5,-2)*{a}
,(10,-1.8)*{b}
,(15,-2)*{c}
,(20,-2)*{d}
,(25,-2)*{e}
,(30,-2)*{f}
,(2,-5)*{a}
,(2,-10)*{b}
,(2,-15)*{c}
,(2,-20)*{d}
,(2,-25)*{e}
,(2,-30)*{f}
,(5,-5)*\cir(#1,0){}
,(10,-5)*\cir(#1,0){}
,(15,-5)*\cir(#1,0){}
,(20,-5)*\cir(#1,0){}
,(25,-5)*\cir(#1,0){}
,(30,-5)*\cir(#1,0){}
,(5,-10)*\cir(#1,0){}
,(10,-10)*\cir(#1,0){}
,(15,-10)*\cir(#1,0){}
,(20,-10)*\cir(#1,0){}
,(25,-10)*\cir(#1,0){}
,(30,-10)*\cir(#1,0){}
,(5,-15)*\cir(#1,0){}
,(10,-15)*\cir(#1,0){}
,(15,-15)*\cir(#1,0){}
,(20,-15)*\cir(#1,0){}
,(25,-15)*\cir(#1,0){}
,(30,-15)*\cir(#1,0){}
,(5,-20)*\cir(#1,0){}
,(10,-20)*\cir(#1,0){}
,(15,-20)*\cir(#1,0){}
,(20,-20)*\cir(#1,0){}
,(25,-20)*\cir(#1,0){}
,(30,-20)*\cir(#1,0){}
,(5,-25)*\cir(#1,0){}
,(10,-25)*\cir(#1,0){}
,(15,-25)*\cir(#1,0){}
,(20,-25)*\cir(#1,0){}
,(25,-25)*\cir(#1,0){}
,(30,-25)*\cir(#1,0){}
,(5,-30)*\cir(#1,0){}
,(10,-30)*\cir(#1,0){}
,(15,-30)*\cir(#1,0){}
,(20,-30)*\cir(#1,0){}
,(25,-30)*\cir(#1,0){}
,(30,-30)*\cir(#1,0){}
#2\end{xy}}}

\newcommand{\double}[1]{\mathbb{#1}}

\newcommand{\cc}{\double{C}}

\newcommand{\rr}{\double{R}}
\newcommand{\zz}{\double{Z}}

\newcommand{\aaa}{\mathcal{A}}

\newcommand{\uu}{\mathcal{U}}

\newcommand{\mm}{\mathcal{M}}

\newcommand{\pp}{\pmatrix}

\newcommand{\dd}{\mathcal{D}}

\newcommand{\llll}{\mathcal{L}}

\newcommand{\mmf}{\hbox{$^f$\hspace{-0.2cm} $\mathcal{M}$}}

\newcommand{\ddf}{\hbox{$^f$\hspace{-0.15cm} $\mathcal{D}$}}

\newcommand{\ttt}{{\rm tr}}

\newcommand{\dee}{\hbox{\rm{D}}}
\newcommand{\de}{\hbox{\rm{d}}}

\newcommand{\ot}{\otimes}
\newcommand{\op}{\oplus}

\newcommand{\bb}{\begin{eqnarray}}
\newcommand{\ee}{\end{eqnarray}}
\newcommand{\eee}{\nonumber\end{eqnarray}}
\newcommand{\qq}{\quad}

\begin{document}

\font\twelve=cmbx10 at 13pt
\font\eightrm=cmr8

\thispagestyle{empty}

\begin{center}

Institut f\"ur Mathematik  $^1$ \\ Universit\"at Potsdam
\\ Am Neuen Palais 10 \\14469 Potsdam \\ Germany\\

\vspace{2cm}

{\Large\textbf{New Scalar Fields in Noncommutative Geometry}} \\

\vspace{1.5cm}

{\large Christoph A. Stephan$^{1,2}$}

\vspace{2cm}

{\large\textbf{Abstract}}
\end{center}
In this publication we present an extension of the Standard Model
within the framework of Connes' noncommutative geometry \cite{con}.
The model presented here is based on a minimal spectral triple \cite{5}
which contains the Standard Model particles, new vectorlike
fermions and a new $U(1)$ gauge subgroup. 
Additionally a new complex scalar field appears that couples
to the right-handed neutrino, the new fermions and the standard 
Higgs particle.

The bosonic part of the action is given by the  spectral
action \cite{con} which also determines relations among the
gauge couplings, the quartic scalar couplings and the Yukawa 
couplings at a cut-off energy of $\sim 10^{17}$ GeV.
We investigate the renormalisation group flow of these relations.
The low energy behaviour allows to constrain the Higgs mass,
the mass of the new scalar and the mixing between these
two scalar fields.

\vspace{2cm}

\noindent
PACS-92: 11.15 Gauge field theories\\
MSC-91: 81T13 Yang-Mills and other gauge theories

\vskip 1truecm

\noindent \\

\vspace{1.5cm}
\noindent $^2$ christophstephan@gmx.de\\

\newpage

\section{Introduction}

We present an extension of the Standard Model in
its noncommutative formulation \cite{con}. This
model is based on the classification of finite
spectral triples \cite{1,2,3,Spinlift,4,5}. It extends
a minimal model found in \cite{5} which contains
the first family of Standard Model fermions as well
as a new family of particles we will call X-particles.
These X-particles are assumed to exist in three
generations, just like the Standard Model particles.
The formulation is done in the recent variant
where the $KO$-dimension  of the internal part
of the  spectral triple is taken to be six \cite{barrett,cc}.

We add to the minimal model right-handed neutrinos
together with their Majorana masses. It turns out
that these right-handed neutrinos open the possibility
to add  Dirac mass terms connecting the right-handed
neutrinos and the left-handed X-particle. These
Dirac mass terms induce through the  fluctuations
of the Dirac operator a new scalar field. This new
field and its interaction with the Higgs field will
be one of the  main concerns of this publication.

Our model has as gauge group $G=U(1)_Y \times
SU(2) \times SU(3) \times U(1)_X$ where the Standard 
Model subgroup $G_{SM}=U(1)_Y \times
SU(2) \times SU(3)$
is broken by the usual Higgs mechanism to $U(1)_{em} \times SU(3)$.
The fate of the new subgroup $U(1)_X$ turns out to 
be closely related to the mass of the X-particles. 
Depending on this mass, the vacuum expectation value
of the new scalar is either zero or nonzero, thus $U(1)_X$
can be broken or remain unbroken.
Both models permit a considerable modification of the Higgs
phenomenology for certain mass regions of the new
scalar particle.  We will explore some of the consequences.
We will focus  on the masses of the Higgs boson
and the new scalar as well as the possible mixing of
the two particles.

Previous attempts to extend the standard model within the
framework of noncommutative geometry proved 
to be extremely difficult. Most of the early attempts 
unfortunately failed to produce physically interesting
models \cite{beyond}. The only known extension which
appear to have an interesting phenomenology just add
new fermions to the standard model  \cite{beyond2,vector}
and possibly new gauge bosons \cite{newcolour}. At least
one of these models, the AC-model, provides for an interesting
dark matter candidate \cite{khlop}. 
But the scalar sector has remained so far 
the usual Higgs sector of the Standard Model.

It would of course also be desirable to understand the origin
of the internal space, i.e. the source of the matrix algebra.
There are hints that a connection to loop quantum gravity
exists \cite{dan}. Also  double Fell bundles
seem a plausible structure in noncommutative geometry \cite{ra1}. 
They could provide a deep connection to category theory
and give better insights into the mathematical structure of
almost-commutative geometries such as the standard model.

Another open problem is the mass mechanism for neutrinos.
In KO-dimension zero the masses are of Dirac type \cite{gracia,neutrino,ra2},
while KO-dimension six also allows for Majorana masses \cite{barrett,cc}
and the SeeSaw mechanism,
although  minor problems  concerning an axiom of
noncommutative geometry may occur \cite{ko6}. Another possibility 
lies in the modification of the spectral action \cite{sit}.

For a numerical analysis of the standard model with SeeSaw mechanism we refer
to \cite{cc,sm1,sm2} for the models with three and four summands
in the matrix algebra. 

This paper is organised as follows: In section two we give the construction
of the internal space based on a minimal Krajewski diagram
found in the classification of finite spectral triples in \cite{5}. This 
diagram contains the first family of the Standard Model fermions
and additionally a new family fermions, the X-particles.
We calculate the lift of the gauge group and
the fluctuated Dirac operator. This fluctuation leads to the 
Standard Model Higgs and a new scalar field.

In the third section we calculate the relevant parts of the
spectral action. This calculation provides the potential
for the Higgs and the  scalar field, as well as constraints
on the quartic couplings, the  Yukawa couplings and 
the gauge couplings of the non-abelian subgroup
of the gauge group. 

The necessary $\beta$-functions needed to evolve
the couplings down to lower energies are given
in section four. 

In section five we analyse the running of the couplings
and the  consequences for the masses of the Higgs
boson and the new scalar. Here we assume that
the mass of the new scalar is roughly of the same
order of magnitude as the Higgs boson mass.
In this analysis we neglect the Dirac mass connecting
the right-handed Neutrino and the left-handed X-particle
and we also ignore the implications of the SeeSaw
mechanism. 

\section{The internal space}
Internal spaces of almost-commutative geometries are
conveniently encoded in Krajewski diagrams \cite{kraj}.
Here we will follow the minimal approach that led to 
a classification of the internal spaces of almost-commutative
geometries \cite{1,2,3,Spinlift,4,5} with respect to the
number of summands in the matrix algebra. In \cite{5}
essentially one model beyond the standard model
results from the classification. 
Picking one of the diagrams leading to this specific model
and extending it as minimally as possible leads
to the model presented here.

To construct the internal space of the model we  
begin by enlarging the minimal Krajewski diagram 2 found
in \cite{5}. In its minimal version this diagram encodes the
first family of the standard model (without a right-handed
neutrino) and a new fermion with Dirac mass term. We will
call this new particle the X-particle. In principle the X-particle
may appear in each family. We add to
this diagram a right-handed neutrino, its Dirac mass term
with the lepton doublet, its Majorana mass term  and
a new Dirac mass term coupling the right-handed neutrino
to the left-handed X-particle. No further mass terms are permitted
by the axioms of noncommutative geometry.

The Krajewski diagram for this model is depicted in figure 1. 
Note that the Majorana mass term does not appear explicitly
since we have left out the antiparticles to keep the
diagram simple. We will not go into the details of
the SeeSaw mechanism following from the Majorana
mass, details can be found in \cite{barrett,cc}.

The model presented here has a minor mathematical
shortcoming which it shares with the almost-commutative
Standard Model if right-handed neutrinos with Majorana
mass are added. 
It turns out \cite{ko6} that the representation of the
right-handed neutrino is incompatible with the axiom
of orientability. But it seems to be necessary to have
such a representation for a working SeeSaw mechanism,
which in turn is required by the constraints put on
the Yukawa-couplings by the spectral action (see section 3).
A model with no such shortcomings  would of course
be desirable.

\begin{figure}
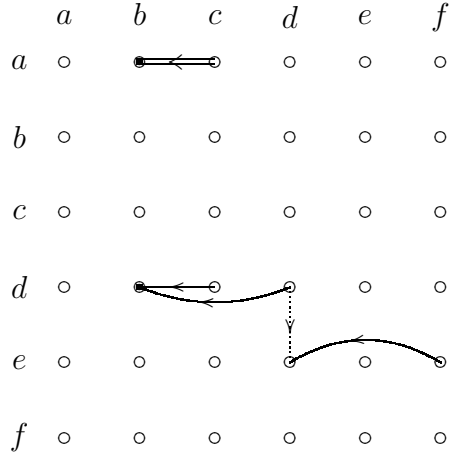

\begin{center}
\begin{tabular}{c}
\rxyn{0.4}{
,(10,-20);(15,-20)**\dir{-}?(.4)*\dir{<}
,(10,-20);(20,-20)**\crv{(15,-22)}?(.4)*\dir{<}
,(10,-20)*\cir(0.2,0){}*\frm{*}
,(10,-5);(15,-5)**\dir2{-}?(.4)*\dir2{<}
,(10,-5)*\cir(0.2,0){}*\frm{*}
,(20,-20);(20,-25)**\dir{..}?(.6)*\dir{>}
,(20,-25);(30,-25)**\crv{(25,-22)}?(.4)*\dir{<}
}
\end{tabular}
\caption{Krajewski diagram of the extended Standard Model. The
dotted  line indicates  the Dirac mass term leading to the new
scalar field $\varphi$.}
\end{center}
\end{figure}

As was shown in \cite{5} the model represented by the
Krajewski diagram in figure 1 allows an anomaly free
lift of the gauge group if the internal algebra is $\aaa = 
\cc \op M_2(\cc) \op M_3(\cc) \op \cc \op \cc \op \cc$.
Here the first four summands are the well known
algebra of the standard model as found in \cite{3,Spinlift,4}.

From the Krajewski diagram  we read off the representation
for $\aaa\ni (a,b,c,d,e,f)$:
\bb 
\rho _L =\pp{b\ot 1_3&0&0 \cr  0& b & 0 \cr 0 & 0& \bar{d} },&&
\rho _R=\pp{c\ot 1_3&0&0&0&0 \cr  0&\bar{c} \ot 1_3&0&0&0
\cr 0&0& \bar{c} &0&0 \cr 0&0&0&\bar{d} &0 \cr 0&0&0&0& f },\cr \cr \cr
\rho _L^c=\pp{1_2 \ot  a&0 &0 \cr  0&d\, 1_2  & 0 \cr 0&0&  e},&&
\rho _R^c=\pp{ a &0&0&0 &0 \cr  0& a &0&0 &0 
\cr 0&0&  d &0 &0 \cr 0&0&0& d &0  \cr 0&0&0& 0&  e },
\ee
 and the Dirac mass matrix:
\bb 
\mm=\pp{M_u\ot 1_3& M_d\ot 1_3 & 0 &0&0\cr 0&  0&M_e &M_\nu &0
\cr 0&  0&0& M_{\nu X} & M_X  },\cr \cr \cr 
\ee
Where 
$ M_u, M_d, M_e, M_\nu \in
M_{2\times
1}(\cc)$ are the usual mass matrices of the quarks and leptons
while $M_{\nu X} \in \cc$ represents the Dirac mass connecting
the right-handed neutrino and the left-handed X-particle and 
$M_X \in \cc$ is the Dirac mass term of the X-particle.

The internal part $\dd$ of the Dirac operator can be decomposed
as follows:
\bb
\dd = \pp{\Delta & M \cr M & \bar{\Delta}}, \quad  {\rm with} \quad
\Delta = \pp{ 0 & \mm \cr \mm^* & 0} 
\ee
The Majorana mass matrix of the right-handed neutrino is
\bb
M = \pp{0&0&0&0 \cr 0& M_M &0&0 \cr 0&0&0&0 \cr 0&0&0&0}, \; \; M_M \in \rr.
\ee
The non-abelian subgroup of unitaries of the matrix algebra $\aaa$ is 
$\uu^{nc} = U(2) \times U(3)$. It contains two $U(1)$ subgroups
via the determinant that
may be lifted to the fermionic Hilbert space \cite{fare}. We will call
these two subgroups suggestively $U(1)_Y$ and $U(1)_X$, since
the first one is nothing else but the Standard Model hypercharge
subgroup and the second one is associated with the X-particles.
The X-particles are neutral with respect to the non-abelian part
of the Standard Model gauge group, i.e. the X-particles
are $SU(2) \times SU(3)$ singlets.

For simplicity we will assume that the hypercharge $U(1)_Y$ couples
only to the Standard Model sector of the model, while the $U(1)_X$ 
couples only to X-particles and the newly emerging scalar field
which we will call $\varphi$. This choice is natural since the
anomaly cancelation forces the Standard Model particles
to couple proportionally to each possible $U(1)$ subgroup of
the gauge group. Therefore the Standard Model only ``sees''
one linear combination of the $U(1)$'s while the X-particles
may see another linear combination. So what we essentially
do by our choice, is setting the electrical charge of the
X-particles to zero.

The anomaly free lift $L$ then decomposes into the usual 
standard model lift $L_{SM}$ which can be found in \cite{fare}
and the lift  $L_X$ acting on the X-particles. This can be written
as
\bb
L(\det(u),\det(v),u,v) = L_{SM}( \det(v), \tilde{u},\tilde{v}) \op L_X(\det(u))
\ee
where $u\in U(2)$, $v \in U(3)$, $\tilde{u} \in SU(2)$ and $\tilde{v} \in SU(3)$.
For $L_{SM}$ we find the standard lift \cite{fare} and for the new part of
the lift $L_X$ we find
\bb
L_X(\det(u)) = {\rm diag}(\det(u)^{Q_X},\det(u)^{Q_X};\det(u)^{-Q_X},\det(u)^{-Q_X}).
\ee
Here $Q_X$ is the charge of the X-particles under $U(1)_X$ and the semicolon
divides the particles from the antiparticles. One notices that the X-particles
couple vectorially to $U(1)_X$ and therefore their Dirac mass $M_X$ is
gauge invariant. 
It follows that the gauge group of our model is 
$G = U(1)_Y \times SU(2) \times SU(3) \times U(1)_X$.

Next we need to fluctuate the Dirac operator \cite{con}  to obtain the gauge 
bosons as well as the Higgs field $\phi$  and the new scalar field 
$\varphi$.
We define the fluctuated Dirac operator $\ddf$ according to \cite{1}:
\bb
\ddf = \sum_i r_i L(\det(u_i),\det(v_i),u_i,v_i) \dd L(\det(u_i),\det(v_i),u_i,v_i)^{-1},
\;\; r_i \in \rr .
\ee
One obtains the standard Higgs doublet $\phi$ embedded into a quaternion
and a new complex scalar field. For definiteness we put the $U(1)_X$ charge of
the X-particles to $Q_X=1$ and therefore the charge of $\varphi$ under
$U(1)_X$ is $Q_X=-1$:
\bb
\ddf|_{\rm rest.}  = M_{\nu X} \, \sum_i r_i \det(u_i)^{-1} = M_{\nu X} \, \varphi, 
\ee
where $\ddf|_{\rm rest.}$ denotes the part of the fluctuated Dirac operator
restricted to the mass matrix that does not commute with the
fluctuation of $L_X$. 

The Majorana mass matrix of the neutrino commutes with
the fluctuation. So we find for the fluctuated mass matrix
\bb
\mmf = \pp{ \phi M_u \ot 1_3 & \phi M_d \ot 1_3 &0&0&0 \cr
0&0& \phi M_e & \phi M_\nu &0 \cr
0&0&0& \varphi M_{\nu X} & M_X }.
\ee
From this mass matrix we can now calculate the spectral action
which will give us the kinetic term of the scalars as well
as the potential for the Higgs field and the new scalar.

\section{Spectral action and constraints on the couplings}

According to \cite{con} the spectral action $S_{CC}$ is given by 
the number of eigenvalues of the Dirac operator $\dee$
up to a cut-off energy $\Lambda$. This can be
written approximately with help of a positive
cut-off function $f$ and then be calculated explicitly
via a heat-kernel expansion:
\bb
S_{CC} = \ttt ( f \left(\frac{\dee^2}{\Lambda^2}\right) ) = \frac{1}{16 \pi^2}
\, \int {\rm dV} ( a_4 f_4 \Lambda^4 + a_2 f_2 \Lambda^2 + a_0 f_0 + o(\Lambda^{-2})
)
\ee
Here $f_i$ are the first moments of the cut-off function $f$. They enter as
free parameters into the model. The heat-kernel coefficients $a_i$ are well
known \cite{gilk} and for the present calculation only $a_2$ and 
$a_0$ will be of concern. Note that we use the numerating convention
of \cite{cc}, where the number of the coefficient $a_i$ corresponds to
the power of $\Lambda$.

The coefficient $a_2$ will give us the mass terms of the potential for
the scalar fields while $a_4$ will provide for the kinetic terms for the 
scalar fields,
the quartic couplings of the potential and also mass terms.
All the following relations
hold at the cut-off energy $\Lambda$. They are not stable
under the renormalisation group flow but they provide for 
the values of the quartic and the Yukawa couplings at the
cut-off energy. From there they have to be evolved down
into the low energy regime using the renormalisation group
equations. 

To calculate the relevant parts of  
$a_2$ and $a_4$ we need the traces of $\ddf^2$ and
$\ddf^4$. Since these calculations get easily quite confusing
we will greatly simplify the matter by putting at this point all negligible mass 
terms to zero. This
will be all quark masses, apart from the top-mass $m_t$, and all
lepton masses apart from the tau-neutrino mass $m_\nu$. We
also only keep the largest Majorana mass $m_M$ of the neutrinos, the
mass $m_X$ of the heaviest X-particle family   and the Dirac mass $m_{\nu X}$
connecting the right-handed tau-neutrino to the heaviest X-particle.

For the traces $\ddf^2$ and $\ddf^4$ we find
\bb
\ttt \, \ddf^2 = 4  [ |\phi|^2 ( 3 m_t^2 + m_\nu^2) + |\varphi|^2 m_{\nu X}^2 +
m_X^2 + \frac{1}{2} m_M^2 ]
\ee
and 
\bb
\ttt \, \ddf^4 &=& 4  [ |\phi|^4 ( 3 m_t^4 + m_\nu^4) + |\varphi|^4 m_{\nu X}^4 +
2 m_\nu^2 m_M^2 |\phi|^2 + 2 m_\nu^2 m_{\nu X}^2 |\phi|^2 |\varphi|^2 
\nonumber \\
&&+2 ( m_M^2 +m_X^2)m_{\nu X}^2 |\varphi|^2+ m_X^4 + \frac{1}{2} m_M^4 ] 
\ee
where $|\cdot|$ is the absolute value including the appropriate trace for
the quaternionic realisation of the Higgs.

From $a_0$ we find the kinetic term  for the scalar fields \cite{con}:
\bb
 \frac{f_0}{2 \pi^2} (3 m_t^2 + m_\nu^2) \ttt((D_\mu \phi)^*(D^\mu \phi))+
 \frac{f_0}{2 \pi^2} m_{\nu X}^2 (D_\mu \varphi)^*(D^\mu \varphi)
\ee
One observes that the scalar fields have mass dimension zero. Therefore
we have to normalise the scalar fields, $\phi \rightarrow \tilde \phi$
and $\varphi \rightarrow \tilde \varphi$  to obtain the standard kinetic
terms of the Lagrangian
\bb
\llll_{kin.}&&=  \frac{f_0}{2 \pi^2} (3 m_t^2 + m_\nu^2) \ttt((D_\mu \phi)^*(D^\mu \phi))+
 \frac{f_0}{2 \pi^2} m_{\nu X}^2 (D_\mu \varphi)^*(D^\mu \varphi) \\
&&=^{\! \! \! \! !}  (D_\mu \tilde \phi)^*(D^\mu \tilde \phi) + (D_\mu \tilde \varphi)^*
(D^\mu \tilde \varphi).
\label{kinetic}
\ee
From this we deduce the normalisation
\bb
|\phi|^2 = \frac{2 \pi^2}{f_0 (3m_t^2 + m_\nu^2)} |\tilde \phi|^2
\quad {\rm and} \quad 
|\varphi|^2 = \frac{2 \pi^2}{f_0 m_{\nu X}^2} |\tilde \varphi|^2,
\label{norm}
\ee
which coincides with the standard normalisation
\bb
\tilde \phi = \frac{1}{\sqrt{2}} \pp{ \phi_1 +i \phi_2 \cr \phi_3+i \phi_4}
\quad {\rm and} \quad 
\tilde \varphi = \frac{1}{\sqrt{2}} ( \varphi_1 +i \varphi_2 )
\ee
for the real scalar fields $\phi_i$ and $\varphi_j$.

Now all the terms that are quadratic in the scalar fields are collected
from $a_2$ and $a_0$ to calculate the mass terms. 
\bb
\llll_{quad} &=& - \frac{f_2}{ \pi^2} (3m_t^2 + m_\nu^2) \Lambda^2 |\phi|^2 
- \frac{f_2}{ \pi^2} m_{\nu x}^2 \Lambda^2 |\varphi|^2
\nonumber \\
&&+ \frac{f_0}{2 \pi^2} m_\nu ^2 m_M^2 |\phi|^2
+ \frac{f_0}{ \pi^2} m_{\nu X}^2 m_X^2 |\varphi|^2
\nonumber \\
&=^{\! \! \! \! !}& -\mu_1^2 |\tilde \phi|^2 -\mu_2^2 |\tilde \varphi|^2
\label{quad}
\ee
leads us with the normalisation (\ref{norm}) to
\bb
\mu_1^2 = 2 \frac{f_2}{f_0} \Lambda^2 -\frac{g_\nu^2}{3 g_t^2 + g_\nu^2} \, m_M^2
\quad {\rm and} \quad 
\mu_2^2= 2 \frac{f_2}{f_0} \Lambda^2 -2 m_X^2,
\ee
where $g_t$ is the Yukawa coupling of the top quark and $g_\nu$
is the Yukawa coupling of the tau neutrino. Here
we have used the fact that $m_i / m_j = g_i / g_j$ where $m_i$ would
be the masses whereas $g_i$ are the corresponding Yukawa couplings.

At this stage we encounter an interesting new phenomenon. It turns out
that the constraints which will be determined later, enforce the cut-off
energy to be $\Lambda \sim 10^{17}$ GeV and the Majorana mass
has been determined to be $m_M \sim 10^{14}$ GeV \cite{sm1,sm2}.
It follows that $\mu_1^2$ is positive and therefore the minimum of 
the Higgs potential is nonzero. For $\mu_2^2$ the situation is
different. Since the mass of the X-particles is gauge-invariant
one would expect it to be of the order of cut-off energy. So depending
on the exact value of $m_X$ the sign of $\mu_2^2$ can be positive
or negative and thus allowing for a nonzero or a zero minimum
of the potential. We will explore these two cases in detail later.

The last term needed is the quartic term of the potential. For the
Lagragian we find:
\bb
\llll_{quart}&=& \frac{f_0}{ \pi^2}\, (3 m_t^4 + m_\nu^4) |\phi|^4 
+ \frac{f_0}{ \pi^2} \, m_{\nu X}^4 |\varphi|^4 
\nonumber \\
&& + \frac{f_0}{ \pi^2} \, m_\nu^2 m_{\nu X}^2 |\phi|^2 |\varphi|^2
\nonumber \\
&=^{\! \! \! \! !}& \frac{\lambda_1}{6} \, |\tilde \phi|^4 +\frac{\lambda_2}{6} \, 
|\tilde \varphi|^4 +\frac{\lambda_3}{3} \, |\tilde \phi|^2 |\tilde \varphi|^2
\label{quart}
\ee
Comparing the coefficients and using the normalistion (\ref{norm})
we obtain the following relations  for  the quartic couplings:
\bb
\lambda_1 = 12 \frac{\pi^2}{f_0} \frac{3 g_t^4 + g_\nu^4}{(3 g_t^2 +g_\nu^2)^2},
\quad \lambda_2 = 12  \frac{\pi^2}{f_0}, 
\quad \lambda_3 = 12 \frac{\pi^2}{f_0} \frac{g_\nu^2}{(3 g_t^2 +g_\nu^2)}
\ee
The last set of relations  to be determined  has its origin in
the fermionic part of the action $S_{ferm}= (\Phi, \dee \Phi)$.
In this case the normalisation (\ref{norm}) leads to 
the identification
\bb
(3m_t^2 + m_\nu^2) |\phi|^2 + m_{\nu X}^2 |\varphi|^2
=^{\! \! \! \! !}
(3g_t^2 + g_\nu^2) |\tilde \phi|^2 + g_{\nu X}^2 |\tilde \varphi|^2
\ee
which gives 
\bb
g_{\nu X}^2 = 3g_t^2 + g_\nu^2 = 2 \, \frac{\pi^2}{f_0}.
\label{yuk}
\ee
At last the cut-off energy is fixed by the relation for the
$SU(2)$ gauge coupling $g_2$ and the $SU(3)$ gauge
coupling $g_3$. At $\Lambda$ the equation
\bb
g_2^2 = g_3^2 = \frac{\pi^2}{2f_0}
\label{gauge}
\ee
has to hold \cite{con}. This allows to eliminate $f_0$ from the
constraints and to combine the previously obtained relations.

Collecting  the conditions for the quartic couplings (\ref{quart}), the Yukawa
couplings (\ref{yuk}) and the gauge couplings (\ref{gauge}) we obtain the final
relations 
\bb
g_2^2 = g_3^2 = \frac{\lambda_1}{24} \frac{(3g_t^2 + g_\nu^2)^2}{3g_t^4 + g_\nu^4} = \frac{\lambda_2}{24} = \frac{\lambda_3}{24} \frac{3g_t^2 + g_\nu^2}{g_\nu^2}
= \frac{1}{4}\, g_{\nu X}^2 = \frac{1}{4}\, (3g_t^2 + g_\nu^2)
\label{relation}
\ee
which are to hold at the cut-off energy $\Lambda$.

\section{The renormalisation group equations}

We will now give  the one-loop $\beta$-functions of the standard model  with $N=3$ generations with X-particles and new scalar field $\tilde \varphi$ to evolve the constraints
 (\ref{relation}) from $E= \Lambda$  down to the low energy regime at $E=m_Z$. We set:
$ t:=\ln (E/m_Z),\qq \de g/\de t=:\beta _g,\qq \kappa :=(4\pi )^{-2}$. 

A mentioned above all fermion masses below the top mass will be neglected. We will
also neglect threshold effects. A Dirac mass $m_\nu$ for the $\tau$  neutrino induced by spontaneous symmetry breaking is admitted and is taken to be 
of the order of the top mass. The  Majorana mass $m_M$ is fixed to
be $\sim 10^{14}$ GeV to obtain the  SeeSaw mechanism \cite {seesaw}. 
The effect of the running of these Majorana masses on the other couplings
 was shown to be tiny \cite{sm1,sm2}, so we will neglect it.
Furthermore the mass $m_X$ of the X-particle will be taken to be of the order
of $\Lambda$. 

Since the Dirac mass $m_{\nu X}$ couples the ultra heavy right-handed neutrino
and left-handed X-particle we will also neglect this coupling by virtue of the
Appelquist-Carazzone decoupling theorem \cite{ac}. 

The gauge couplings for the subgroups of  the gauge group 
$G= U(1)_Y \times SU(2) \times SU(3) \times U(1)_X$ are denoted
$g_1$, $g_2$, $g_3$ and $g_4$.
The $\beta$-functions are \cite{mv,jones}:
\bb 
\beta _{g_i}&=&\kappa b_ig_i^3,\qq b_i=
{\textstyle
\left( \frac{20}{9} N+\frac{1}{6},-\frac{22}{3}+\frac{4}{3} N+\frac{1} {6},
-11+\frac{4}{3} N,  \frac{1}{3}\right) },
\\ \cr
\beta _t&=&\kappa
\left[ -\sum_i c_i^ug_i^2 + \,\frac{9}{2}\,g_t^2
\,\right] g_t, \quad c_i^t=\left( \frac{17}{12},\frac{9}{4} , 8 ,0\right), \\
\beta _{\lambda_1} &=&\kappa
\left[ \,\frac{9}{4}\,\left( g_1^4+2g_1^2g_2^2+3g_2^4\right)
-\left( 3g_1^2+9g_2^2\right) \lambda_1
+12 g_t^2 \lambda_1 -36 g_t^4 +4\lambda_1^2 +\frac{2}{3} \lambda_3^2 \right] ,\\
\beta _{\lambda_2} &=&\kappa
\left[ \, 36 g_4^4 - g_4^2  \lambda_2 + \frac{10}{3} \lambda_2^2 +\frac{4}{3} \lambda_3^2 \right] ,\\
\beta _{\lambda_3} &=&\kappa
\left[ \, -\left( \frac{3}{2} g_1^2+\frac{9}{2} g_2^2+ 6 g_4^2 \right) \lambda_3
+6 g_t^2 \lambda_3 +\frac{4}{3}\lambda_3^2 +2 \lambda_1 \lambda_3
+\frac{4}{3} \lambda_2 \lambda_3 \right] ,
\label{renorm}
\ee
The four gauge couplings decouple from the other equations 
\bb 
g_i(t)=g_{i0}/\sqrt{1-2\kappa b_ig_{i0}^2t}.
\ee
The initial conditions are taken from experiment \cite{data}:
\bb 
g_{10}= 0.3575,\qq
g_{20}=0.6514,\qq
g_{30}=1.221.
\ee
Since $g_1$ is unconstrained the unification  scale $\Lambda $ is the solution of 
$g_2(\ln (\Lambda /m_Z))=g_3(\ln  (\Lambda /m_Z))$,
\bb 
\Lambda = m_Z\exp\frac{g_{20}^{-2}-g_{30}^{-2}}{2\kappa (b_2-b_3)} \,=\,1.1\times 10^{17}\  {\rm GeV},
\ee
and is independent of the number of generations.
Next we choose $g_\nu=Rg_t$ at $E=\Lambda$ in order to recover
the correct top quark mass. Then we  solve  numerically the evolution equations for $\lambda_1, \ \lambda_2, \ \lambda_3$ and $g_t$ 
with initial conditions at $E=\Lambda$ from the noncommutative  constraints 
(\ref{relation}):
\bb 
g_2^2 = \frac{\lambda_1}{24} \frac{(3 + R^2)^2}{3 + R^4} = \frac{\lambda_2}{24} = \frac{\lambda_3}{24} \frac{3 + R^2}{R^2}
= \frac{3 + R^2}{4}\,g_t^2 \,.
\ee
We note that these constraints imply that all couplings remain  perturbative and
at our energies we obtain the pole masses of the Higgs, the new scalar field 
and the top quark. The top quark mass is then given by 
\bb
m_t=\sqrt{2}\,\frac{g_t(m_t)}{g_2(m_t)}\,m_W,
\ee
while the Higgs mass and the mass of the new particle are obtained by
diagonalising the possibly non-diagonal mass matrix generated by the $\lambda_3$ 
coupling term.
The parameter $R$ will be of no further interest to us here. It will be 
fixed to $R\sim 1.5$ and allows to recover the top mass $m_t \sim 170$ GeV. 
It turns out to be rather insensitive to the running of the non-Standard Model
couplings.

\section{Physical consequences}

In the following we will replace $\tilde \phi$ and $\tilde \varphi$ 
by $\phi$ and $\varphi$ to obtain a simpler notation. This 
is not to be confused with the $\phi$ and $\varphi$ which 
had been normalised in (\ref{norm}).

We will now examine the basic physical features of the model
presented above. Since the mass term $\mu_2^2$ of the
new scalar field $\varphi$ in the quadratic part of the Lagrangian
(\ref{quad}) may have either positive or negative sign, depending
on the X-particles mass term $m_X$, we  treat these
cases separately.

The new scalar field $\varphi$ and the new gauge coupling $g_4$
associated to the gauge subgroup $U(1)_X$ do not influence 
the running  of the  non-abelian gauge couplings $g_2$ and
$g_3$. Therefore the cut-off energy $\Lambda$ which is determined
by the constraint $g_2=g_3$, valid at $\Lambda$, remains
unaltered compared to the pure Standard Model value of 
$\Lambda = 1.1 \times 10^{17}$ GeV \cite{cc}. 

Our main focus will be on the masses of the Higgs particle $\phi$
and the  new scalar field $\varphi$. Putting together the
relevant Lagrangians (\ref{quad}) and (\ref{quart}) we get 
the potential
\bb
V(\phi,\varphi) =  -\mu_1^2 |\tilde \phi|^2 -\mu_2^2 |\tilde \varphi|^2 +
\frac{\lambda_1}{6} \, |\tilde \phi|^4 +\frac{\lambda_2}{6} \, 
|\tilde \varphi|^4 +\frac{\lambda_3}{3} \, |\tilde \phi|^2 |\tilde \varphi|^2.
\label{pot}
\ee

From the constraints (\ref{relation}) we get that $\lambda_i(\Lambda)> 0$
for $i=1..3$. To ensure that the potential is bounded from below, we will require that
the quartic couplings remain positive under the  renormalisation group flow.
This requirement will put a limit on the possible values of $g_4$. 

\subsection{The case $\mu_2 < 0$}

We remind the reader that the quadratic coupling $\mu_2^2$ of the scalar field
$\varphi$ is given by
\bb
\mu_2^2 = 2 \frac{f_2}{f_0} \, \Lambda - 2 m_X^2.
\ee
Since the Majorana mass $m_M$ of the neutrino is of the order
of $\sim 10^{14}$ GeV the quadratic coupling $\mu_1^2$ of the 
Higgs field $\phi$ has always positive sign. 
Following \cite{cc}
we can estimate  $f_2/f_0 \sim \tau^2 / \rho^2$. Here
$\tau \sim 5.1$ and $\rho^2= G(\Lambda) / G(M_Z)$ 
is a measure for the running of Newton's constant $G$.
Assuming  a moderate running
of $G$ at high energies and   $m_X \gtrsim (f_2/f_0)\, \Lambda$ it is possible to 
achieve $\mu_2 < 0$.

As a consequence of $\mu_2<0$ the potential (\ref{pot}) of the
scalar fields implies a zero vacuum expectation value (vev)
$|\langle \hat \varphi \rangle| = 0$ for the 
new scalar field. Therefore the vev of the 
Higgs field is $|\langle \hat \phi \rangle| = v_1/\sqrt{2} = 
\sqrt{6 \mu_1^2 / \lambda_1}/\sqrt{2}$.

As a further consequence of the zero vev of $\varphi$ the gauge group $U(1)_X$
remains unbroken and the combined scalar sector breaks the
whole gauge group as
\bb
U(1)_Y \times SU(2)\times SU(3) \times U(1)_X \longrightarrow
U(1)_{em} \times SU(3) \times U(1)_X \, .
\ee

Furthermore the $\varphi$ is not charged under the Standard Model
gauge subgroup while the Higgs field $\phi$ is uncharged under
$U(1)_X$. It follows that the gauge bosons do not mix, i.e. the
mass of the $W^\pm$-boson is still given by $m_{W^{\pm}}=(g_2/2) \, v_1$.
Therefore $v_1$ takes its experimental value $v_1 = 246$ GeV.
This will remain true even if the vev of the new scalar field
is nonzero.
In the case of $|\langle \hat \varphi \rangle| = 0$ the $U(1)_X$ gauge boson 
remains massless 
and appears as a second photon $\gamma '$.

If the vev of $\varphi$ is zero we have for the Higgs mass 
\bb
m_H^2 = \frac{4}{3} \, \frac{\lambda_1(m_H)}{g_2(m_Z)^2} \, m_W^2,
\label{higgs}
\ee
while the mass of the new scalar field is given by
\bb
m_\varphi^2 = \frac{2}{3} \, \frac{\lambda_3(m_{\varphi}}{g_2(m_Z)^2} \, m_W^2
+ \mu_2(m_\varphi)^2.
\label{phi}
\ee
The parameters which determines  the Higgs mass is $\lambda_1$. 
The free parameters which determine  the mass
of the new scalar are  $\mu_2$ and implicitly through
the $\beta$-functions of the renormalisation group equations, $g_4$.

We pursue now the following general strategy: First we evolve $g_2$ to
the cut-off energy $\Lambda = 1.1 \times 10^{17}$ GeV. Using
the constraints (\ref{relation}) the quartic couplings $\lambda_1$, $\lambda_2$,
$\lambda_3$ as well as the top quark Yukawa coupling $g_t$ and
the parameter $R$ for the right-handed neutrino are fixed.
Then $g_4$ and $\mu_2$ are chosen at $m_Z$ as a free parameters .

Having fixed the free couplings we use the renormalisation group
equations (\ref{renorm}) to evolve the couplings down to low
energies. When the pole masses have been reached we
calculate the mass of the physical Higgs boson using (\ref{higgs})
and the mass of the new scalar using (\ref{phi}). For simplicity
we will only consider the region where $m_H/2 \leq m_\varphi \leq 500$ GeV.

The initial conditions are taken from experiment \cite{data}:
\bb 
g_{10}= 0.3575,\qq
g_{20}=0.6514,\qq
g_{30}=1.221.
\ee
For the top quark mass we take $m_t = 170$ GeV
and for the $W^\pm$ boson mass $m_W = 80.4$ GeV. We will ignore
all the uncertainties on these values since we are only interested in the general
behaviour of the model. A detailed investigation using latest 
data will follow in a later publication.

To ensure that $\lambda_2$ remains positive throughout the 
running of the couplings we have to take $g_4(m_Z) \leq 0.845$.
The quartic Higgs coupling turns out to be almost unaltered
by any choice of $\mu_2$ and $g_4$ within the range specified
above.

Let us first study the most interesting case where $\mu_2(m_\varphi) \ll m_W$.
Assuming this we obtain 
\bb
m_\varphi^2 \approx \frac{2}{3} \, \frac{\lambda_3(m_{\varphi}}{g_2(m_Z)^2} \, m_W^2.
\ee
In figure 2 we have plotted  $\lambda_2(m_\phi)$ and
$\lambda_3(m_\phi)$ with respect to $g_4(m_Z)$. One observes 
the steep drop of $\lambda_2$ as $g_4$ reaches its critical value of 
$0.845$.
\begin{figure}
\begin{center}
\psfrag{AA}[Bl][l][1][0]{ $g_4$}
\psfrag{BB}[][l][1][90]{ $\lambda_{2/3}$}
\includegraphics[scale=0.6]{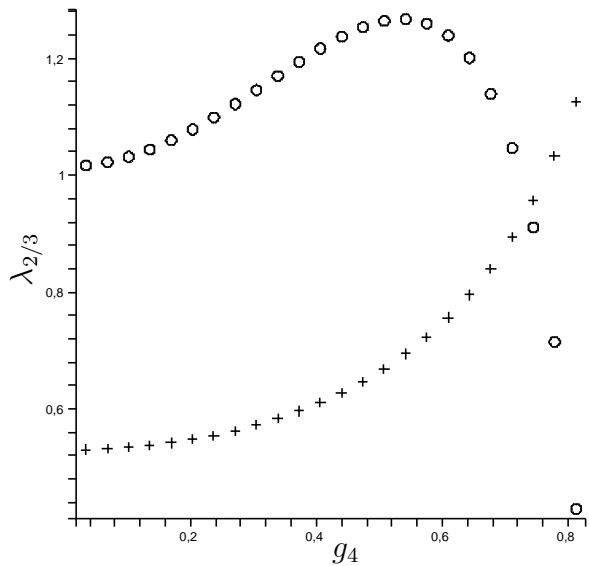}
\caption{Dependence of $\lambda_2(m_\varphi)$ (circles) and $\lambda_3(m_\varphi)$ 
(crosses) on $g_4(m_Z)$ for $\mu_2(m_\varphi) \ll m_W$.}
\end{center}
\end{figure}
\begin{figure}
\begin{center}
\psfrag{AA}[Bl][l][1][0]{ $g_4$}
\psfrag{BB}[][l][1][90]{ $m/$ GeV}
\includegraphics[scale=0.6]{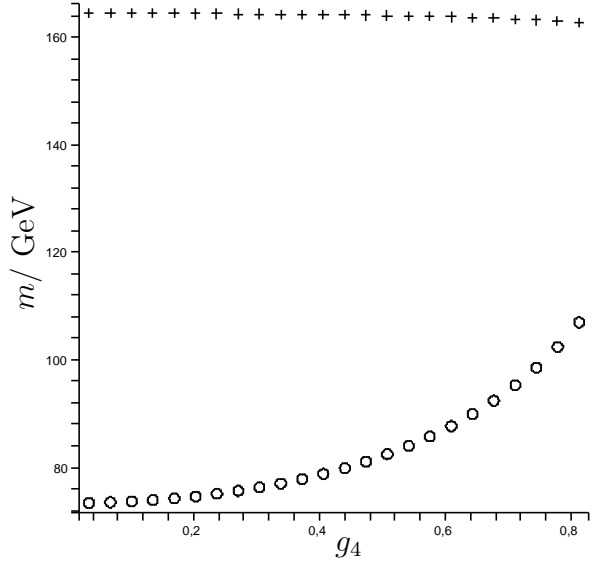}
\caption{Dependence of the Higgs mass $m_H$ (circles) and the
mass of the new scalar $m_\varphi$ 
(crosses) on $g_4(m_Z)$ for $\mu_2(m_\varphi) \ll m_W$.}
\end{center}
\end{figure}

In figure 3 we have plotted the Higgs mass $m_H$ and the mass
of the new scalar $m_\varphi$ with respect to $g_4(m_Z)$.
Indeed the Higgs mass is almost independent of $g_4$ and
takes its pole mass at $m_H \approx 163$ GeV. Compared to
the Standard Model value of $\sim 168$ GeV this is only a minor
decrease. In contrast to that, the mass of the new scalar field depends 
rather strongly on $g_4$. It reaches from $m_\varphi \approx 73$ GeV
for $g_4< 0.1$ to $m_\varphi \approx 107$ GeV for $g_4 \approx 0.81$.

If $\mu_2$ is increased and becomes comparable to $m_W$ it will
merely shift the mass of $m_\varphi$ upwards as can be seen
from (\ref{phi}).

Physically the most interesting case is certainly $2 m_\varphi \leq m_H $.
In this case the Higgs boson may decay into the new scalar fields. But
these scalar fields do not couple to Standard Model fermions and
would thus be unobservable in particle detectors used at Tevatron
or LHC. The decay width of the Standard Model Higgs into $W^\pm$
bosons will therefore be decreased. This fact could reconcile the
predicted Higgs mass of $m_H \approx 163$ GeV with recent
Tevatron data \cite{Tevatron}.

\subsection{The case $\mu_2>0$}

Let us now turn to the case $\mu_2> 0$, i.e.  $m_X < (f_2/f_0)\, \Lambda$.
Now the potential $V(\phi,\varphi)$ (\ref{pot}) requires both 
scalar fields to have nonzero vevs, $|\langle \hat \phi \rangle| = v_1/\sqrt{2} \neq  0$ and
$|\langle \hat \varphi \rangle| = v_2/\sqrt{2} \neq 0$. It is still possible to 
determine the  vev $v_1$ of
the Higgs field since the relation for the $W^\pm$ boson mass, 
$m_{W^{\pm}}=(g_2/2) \, v_1$, continues to hold. 
The vev of $\varphi$ in contrast is a free parameter, essentially
determined by $\mu_2$. 

We obtain the physical real Higgs $h_1$ and real scalar field $h_2$ in the 
standard notation
\bb
\phi = \frac{1}{\sqrt{2}} \, \pp{0 \cr h_1 + v_1}, \quad {\rm and}
\quad \varphi = \frac{1}{\sqrt{2}} \, (h_2 +v_2)
\ee 
But now the Higgs field and the new scalar mix through
the $\lambda_3|\phi|^2 |\varphi|^2$ term in the potential and
the two nonzero vevs produce a non-diagonal mass matrix.
The physical scalar fields correspond therefore to the
mass eigenvalues which are easily calculated to be \cite{Emam}
\bb
m_{H_1,H_2} = \frac{\lambda_1}{6} \, v_1^2 + \frac{\lambda_2}{6} \, v_2^2
\mp \sqrt{ \left(\frac{\lambda_1}{6} \, v_1^2 - \frac{\lambda_2}{6} \, v_2^2 \right)^2
+ \frac{\lambda_3^2}{9} \, v_1^2 v_2^2},
\ee
where the real mass eigenstates $H_1$ and $H_2$ are given by 
\bb
\pp{H_1 \cr H_2} = \pp{\cos \theta & - \sin \theta \cr \sin \theta & \cos \theta} 
\pp{h_1 \cr h_2}
\ee
and
\bb
\tan (2 \theta) = \frac{ 2 \lambda_3 v_1 v_2}{\lambda_1 v_1^2 - \lambda_2 v_2^2}\, .
\ee
\begin{figure}
\begin{center}
\psfrag{AA}[Bl][l][0.5][0]{ $v_2/$ GeV}
\psfrag{BB}[][l][0.5][90]{ $m/$ GeV}
\includegraphics[scale=0.5]{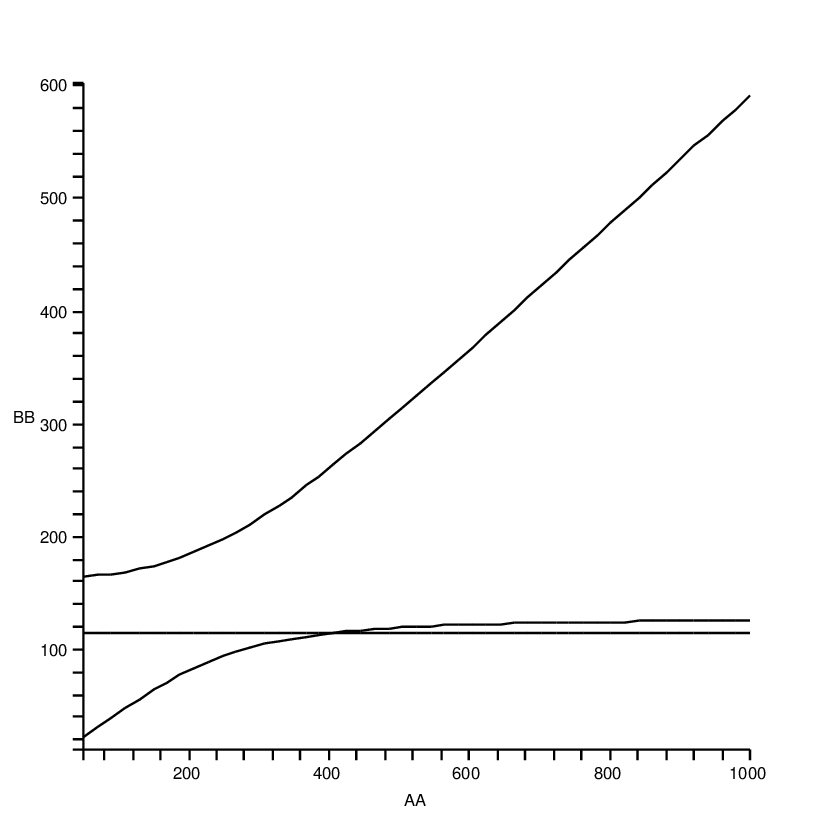}
\psfrag{AA}[Bl][l][0.5][0]{ $v_2/$ GeV}
\psfrag{BB}[Bl][l][0.6][0]{ $\theta$}
\hskip0.5cm
\includegraphics[scale=0.5]{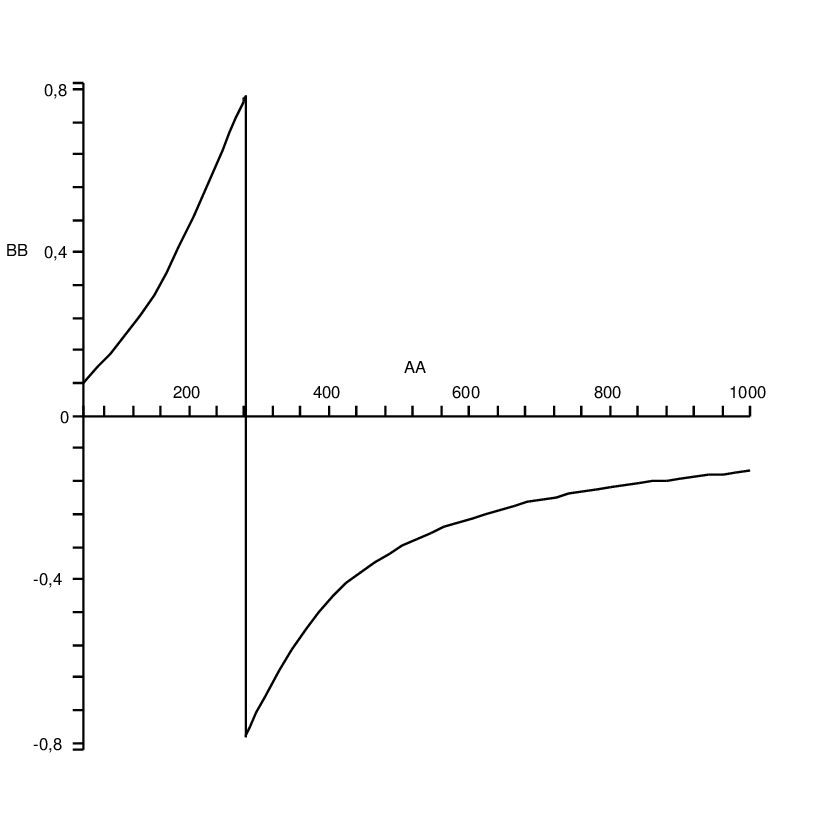}
\caption{$g_4(m_Z)=0.01$: The left figure shows the dependence of
$m_{H_1}$ and $m_{H_2}$ on $v_2$. 
The constant line is the 114 GeV experimental LEP threshold
for the Standard Model Higgs mass. The right figure shows
the dependence of the mixing angle $\theta$ on $v_2$.}
\end{center}
\end{figure}
\begin{figure}
\begin{center}
\psfrag{AA}[Bl][l][0.5][0]{ $v_2/$ GeV}
\psfrag{BB}[][l][0.5][90]{ $m/$ GeV}
\includegraphics[scale=0.5]{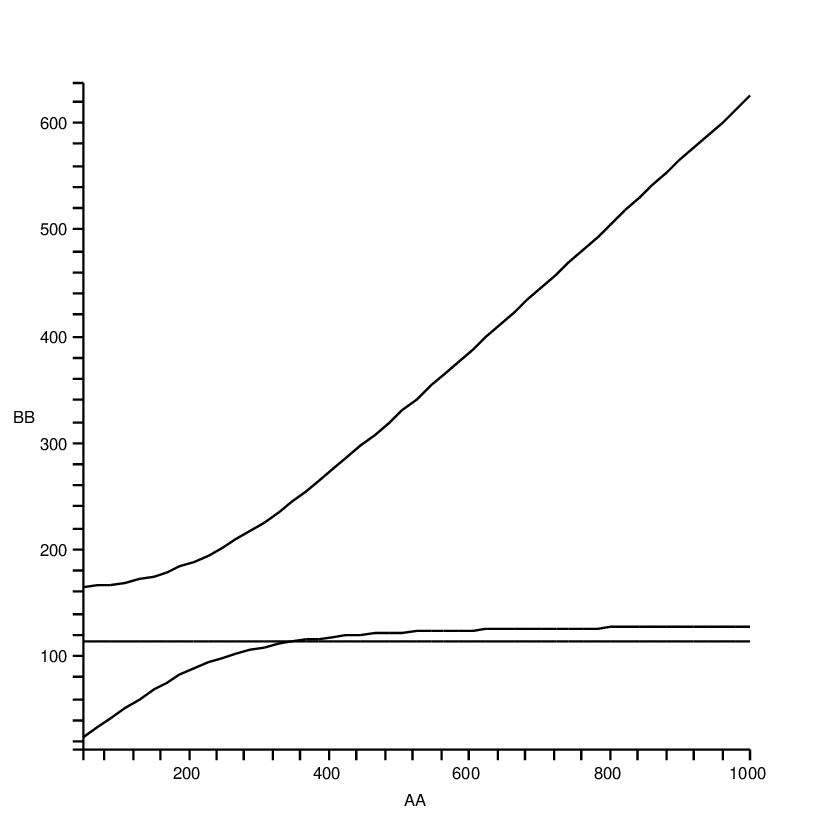}
\psfrag{AA}[Bl][l][0.5][0]{ $v_2/$ GeV}
\psfrag{BB}[Bl][l][0.6][0]{ $\theta$}
\hskip0.5cm
\includegraphics[scale=0.5]{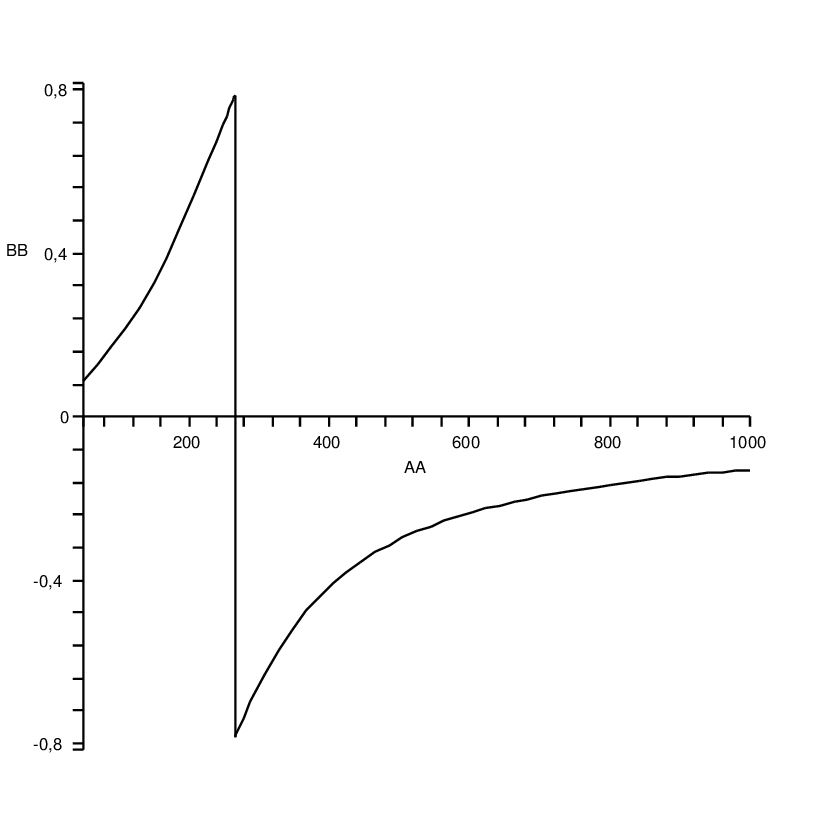}
\caption{$g_4(m_Z)=0.3$: The left figure shows the dependence of
$m_{H_1}$ and $m_{H_2}$ on $v_2$. 
The constant line is the 114 GeV experimental LEP threshold
for the Standard Model Higgs mass. The right figure shows
the dependence of the mixing angle $\theta$ on $v_2$.}
\end{center}
\end{figure}
The mass eigenvalues as well as the mixing angles of $H_1$ and
$H_2$ depend on the $U(1)_X$ gauge coupling $g_4$. For comparison
we have plotted the two mass eigenvalues $m_{H_1}$ and
$m_{H_2}$ and the mixing angle $\theta$ in dependence of $v_2$ 
for the three values $g_4(m_Z)=0.01$ (figure 4), $g_4(m_Z)=0.3$ (figure 5)
and $g_4(m_Z)=0.7$ (figure 6).
\begin{figure}
\begin{center}
\psfrag{AA}[Bl][l][0.5][0]{ $v_2/$ GeV}
\psfrag{BB}[][l][0.5][90]{ $m/$ GeV}
\includegraphics[scale=0.5]{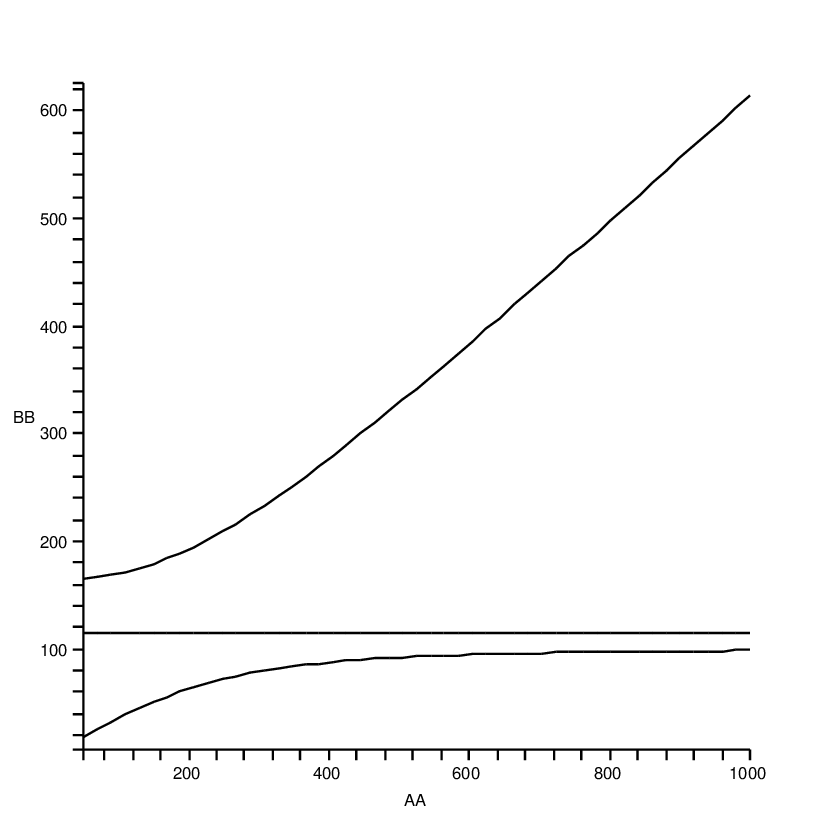}
\psfrag{AA}[Bl][l][0.5][0]{ $v_2/$ GeV}
\psfrag{BB}[Bl][l][0.6][0]{ $\theta$}
\hskip0.5cm
\includegraphics[scale=0.5]{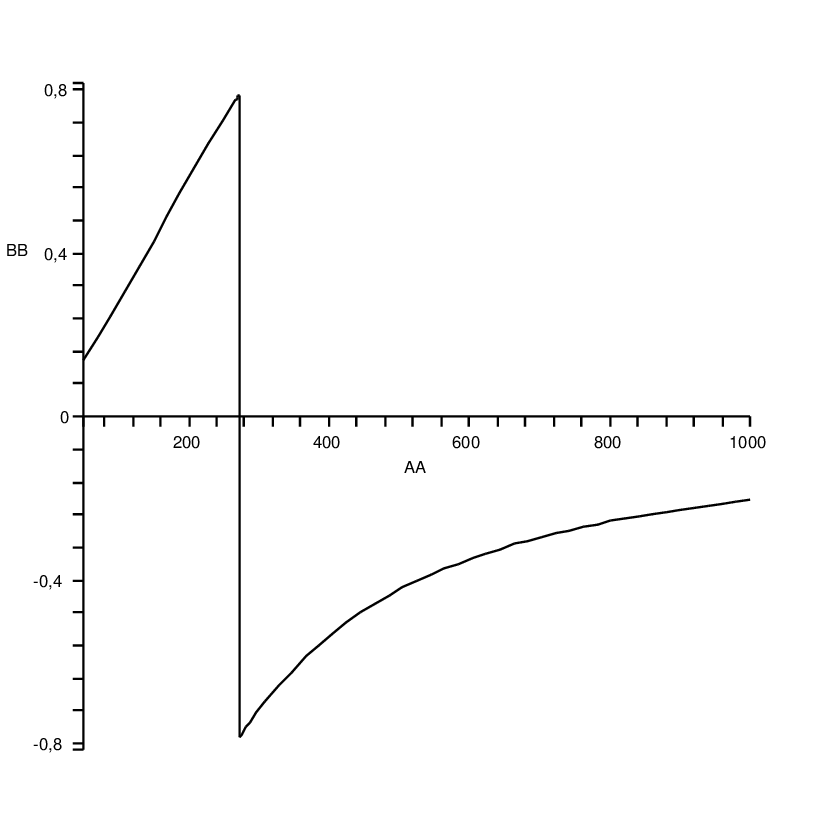}
\caption{$g_4(m_Z)=0.7$: The left figure shows the dependence of
$m_{H_1}$ and $m_{H_2}$ on $v_2$. 
The constant line is the 114 GeV experimental LEP threshold
for the Standard Model Higgs mass. The right figure shows
the dependence of the mixing angle $\theta$ on $v_2$.}
\end{center}
\end{figure}
The smaller mass eigenvalue approaches in each case 
a maximal value as $v_2$ becomes big. This maximal
mass crosses the  $\sim 114$ GeV line for $g_4  \lesssim 0.64$.
But  throughout the whole range of $v_2$ the mixing angles remain 
strictly nonzero.
Therefore neither mass eigenstate corresponds exactly to
the Standard Model model Higgs.

Since the  vev of the new scalar field is now nonzero the
scalar sector breaks the gauge group as follows
\bb
U(1)_Y \times SU(2)\times SU(3) \times U(1)_X \longrightarrow
U(1)_{em} \times SU(3)  \, ,
\ee
where the gauge group $U(1)_X$ is broken into the
discrete group $\zz_2$.
As a consequence of the breaking of the gauge group 
the $U(1)_X$ gauge boson acquires a mass which
is determined by the vev $v_2$ of the new scalar $\varphi$
and the gauge coupling $g_4$.
One finds for the mass of the  $Z'$ boson
\bb
m_{Z'} = g_4 \, v_2.
\ee

\section{Conclusion and outlook}

In this publication we presented an extension  of the
Standard Model within the framework of Connes' noncommutative
geometry  \cite{con}. To obtain the model
we slightly extended an extension of the Standard Model
found in the classification of minimal spectral triples \cite{5}.
The fermionic sector of this  minimal spectral triple contains 
the first family of the Standard Model fermions and an
extra particle which we call the X-particle. This minimal
model allows for an anomaly free charge assignment under
the enlarged Standard Model gauge group $G=U(1)_Y \times
SU(2) \times SU(3) \times U(1)_X$. We assumed that
the Standard Model particles, including the Higgs doublet,
are neutral to the new $U(1)_X$ gauge group, while 
the X-particles are neutral to the Standard Model gauge group
but couple vectorially to $U(1)_X$. Consequently their
masses are gauge invariant and are therefore assumed
to be of the order of the cut-off energy $\Lambda \sim 10^{17}$ GeV.

To this basic model we add right-handed neutrinos, together
with their Majorana mass terms. At this stage something 
interesting happens. The axioms of noncommutative 
geometry, which can be encoded in  Krajewski
diagrams \cite{kraj}, permit an additional Dirac mass term.
This new mass term connects the  right-handed neutrinos
and the left-handed X-particles. Fluctuating the Dirac operator
with the lifted group of unitaries of the internal matrix algebra
$\aaa = 
\cc \op M_2(\cc) \op M_3(\cc) \op \cc \op \cc \op \cc$
then produces the standard Higgs field and a new scaler
field. Calculating the spectral action for this model results
in a term mixing these two fields, thus altering the Standard
Model Higgs sector considerably.

An intriguing fact of the spectral action principle is that it
allows to fix the quartic couplings of the model at a cut-off
energy. This property
has been exploited to calculate the value of the coupling
constants at low energies. From these values the
masses of the Higgs field and its coupling to the new scalar
can be calculated.

It turns out that the sign of the quadratic coupling of the new
scalar field is determined by the mass of the X-particles.
If at least one family of X-particles is sufficiently heavy
compared to the cut-off energy,
the sign is negative and we have a mass term. If the
mass is small compared to the cut-off energy we 
obtain a positive sign and the new scalar field acquires
a nonzero vacuum expectation value.

These two cases have been studied and the
renormalisation group technique has been applied
to the quartic couplings. The results have been 
presented in section 5. We found that the numerical
values depend on the $U(1)_X$ gauge coupling 
$g_4$ as well as the mass of the X-particles which
enters implicitly through the quadratic coupling 
of the new scalar.

The phenomenology of this model seems intriguing.
Since the classical prediction of the Higgs mass
of $m_H \approx 170$ GeV \cite{con,cc}
from the spectral action is almost certainly excluded
by the Tevatron \cite{Tevatron} the model
presented here may open a new window.

For the case of a zero vacuum expectation value
the mass of the Higgs particle remains almost unchanged
compared
to the Standard Model value of $m_H \approx 170$ GeV.
But the mass of the new particle can be as low 
as $m_\varphi \approx 73$ GeV which is less than
half the Higgs mass. Therefore the Higgs may decay
into the new scalar thus changing its decay width.
This could perhaps evade the restrictions posed
by the Tevatron \cite{Tevatron}.

For the case of a nonzero vacuum expectation value
the new scalar and the Higgs mix considerably.
The mass eigenstates will in general consist of
a rather light scalar particle $m_{H_1} \sim 120$ GeV
and a heavy particle $m_{H_2} \geq 170$ GeV.

Similar models with additional real and complex 
scalar fields have been studied before. For
example the so called stealth model \cite{Bij} where the new scalar
field can hide the Higgs field completely from detection. This model
might also provide an interesting candidate for dark matter \cite{Burgess}.
See also \cite{Barger} where a closely related model has been
studied, the main difference to our model being that the new
$U(1)$ group is assumed to be a global symmetry. Models
with gauged new $U(1)$ group have also been considered, see
\cite{Emam} for a $(B-L)$-type extension of the Standard Model.

A detailed study of the phenomenology of the model presented
here is in progress and will be published soon. 
Open issues are the compatibility with LEP data and Tevatron
data, the existence of viable dark matter candidates and
perhaps a mechanism to obtain the baryon asymmetry of the
Universe. If these problems could be (partially) solved by
the model presented here, this would be a rather strong
case for the spectral action principle. It is intriguing that 
despite the
new degrees of freedom like the gauge coupling $g_4$
and the mass of the X-particles, the resulting models
are still extremely constrained by the relations among
the couplings (\ref{relation}) at the cut-off energy.

\subsection*{Acknowledgements}

The author gratefully acknowledges the funding of his work
by the Deutsche Forschungsgemeinschaft.

\vfil\eject
\enlargethispage{1cm}
\thispagestyle{empty}

\end{document}